\documentclass[%
superscriptaddress,
reprint,
showpacs,
 amsmath,
 amssymb,
 aps,
pra,
]{revtex4-1}

\usepackage{graphicx}
\usepackage{dcolumn}
\usepackage{bm}
\usepackage{amsmath}
\usepackage{amssymb}
\usepackage{xcolor}

\begin{document}

\title{
Shaped nozzles for cryogenic buffer gas beam sources
}

\author{Di Xiao}
\affiliation{Department of Physics, University of Nevada, Reno NV 89557, USA}
\author{David M. Lancaster} 
\affiliation{Department of Physics, University of Nevada, Reno NV 89557, USA}
\author{Cameron H. Allen} 
\affiliation{Department of Physics, University of Nevada, Reno NV 89557, USA}
\author{Mckenzie J. Taylor} 
\affiliation{Department of Physics, University of Nevada, Reno NV 89557, USA}
\author{Thomas A. Lancaster} 
\affiliation{Department of Physics, University of Nevada, Reno NV 89557, USA}
\affiliation{Department of Biomedical Engineering, University of Michigan, Ann Arbor MI 48109, USA}
\author{Gage Shaw}
\affiliation{Department of Physics, University of Nevada, Reno NV 89557, USA}
\author{Nicholas R. Hutzler}
\affiliation{Division of Physics, Mathematics, and Astronomy, California Institute of Technology, Pasadena CA 91125, USA}
\author{Jonathan D. Weinstein} 
\email{weinstein@physics.unr.edu}
\homepage{http://www.physics.unr.edu/xap/}
\affiliation{Department of Physics, University of Nevada, Reno NV 89557, USA}



\begin{abstract}
Cryogenic buffer gas beams are important sources of cold molecules. In this work we explore the use of a converging-diverging nozzle with a buffer-gas beam. We find that, under appropriate circumstances, the use of a nozzle can produce a beam with improved collimation, lower transverse temperatures, and higher fluxes per solid angle.
\end{abstract}

\pacs{XXX}



\maketitle

\section{Introduction}

The cryogenic buffer-gas beam  (CBGB) has become a powerful tool in the field of cold molecules, delivering high fluxes of both stable molecules and molecular radicals \cite{maxwell2005high, patterson2007bright, van2009electrostatic, lu2011cold, barry2011bright, hutzler2011cryogenic, hutzler2012buffer, patterson2014slow}. 
It has seen use in a variety of experiments, including laser cooling \cite{shuman2010laser, zhelyazkova2014laser, chae2017one, 1808.01067}, precision measurement \cite{baron2013order}, high-resolution spectroscopy \cite{iwata2017high},  and has been proposed for use as a source of cold molecules for Stark deceleration and trapping \cite{fabrikant2014method}, as well as studies of cold chemistry.

In a CBGB source, molecules are produced inside 
a cryogenic cell filled with a high density of buffer gas. Both molecules and buffer gas are extracted through an aperture into a vacuum chamber
\cite{hutzler2012buffer}.
In this paper, we investigate replacing the cell aperture with a  de Laval (converging-diverging) nozzle, and measure its effects on both molecular flux and the properties of the beam of molecules produced.

Nozzle shape has been shown to be important in other cold molecular beam sources.
%
%
In pulsed supersonic beams, the use of a nozzle has been shown to play a significant role in the angular distribution of molecules produced \cite{luria2011generation}; a diverging nozzle produces a significantly more collimated distribution than a simple aperture.
The CRESU technique for generating cold molecular beams relies on the expansion of gas through de Laval nozzles to both cool the molecular beam and to produce a highly collimated beam \cite{fournier2017low}.
In a very different application, de Laval nozzles have a long history of use in rocketry for maximizing thrust by minimizing the temperature and transverse spread of the rocket exhaust \cite{sutton2016rocket}.

With a CBGB source, recent work has shown that shaping the interior of the cell's exit aperture (before the nozzle ``throat'', the location of minimum diameter) can provide an improvement in molecular flux \cite{singh2018optimized}, consistent with prior work investigating the influence of flow dynamics inside the cell on the flux of extracted molecules \cite{bulleid2013characterization}.
Moreover, most prior work with CBGB sources (without the use of a converging-diverging nozzle) have shown significant expansion cooling of the beam at high buffer-gas flows. This cooling has generally resulted in low rotational temperatures and narrow axial velocity distributions, but the transverse velocity spread is typically higher than the axial \cite{barry2011bright,hutzler2011cryogenic,hutzler2012buffer}. 
Our hope is that by using a de Laval nozzle we can reduce the transverse velocity spread, resulting in a reduction in the angular spread of the CBGB. 
If this can be achieved without a reduction in the total number of molecules extracted from the cell, it would result in higher fluxes of molecules per solid angle, an important figure of merit for cold molecular beams.


\subsection{Ideal cooling}

\begin{figure}[ht]
    \begin{center}
      \includegraphics[width=.75\linewidth]{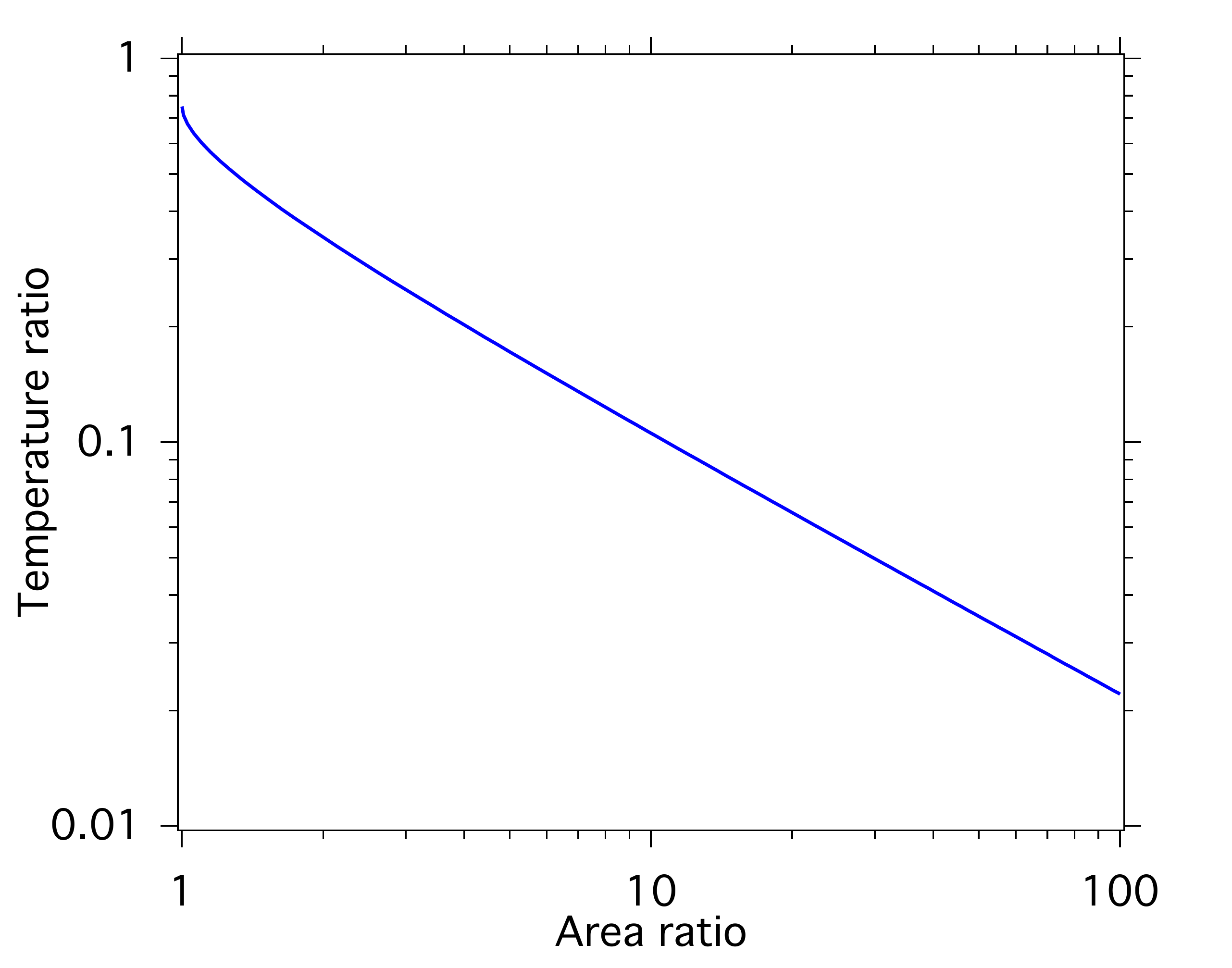}
    \caption{
    \label{fig:cooling} 
    Ideal cooling for a monatomic gas exiting a converging-diverging nozzle as a function of the ratio of the nozzle's exit area to its throat area, calculated as per reference \cite{sutton2016rocket}.
	}     
    \end{center}
\end{figure}

Within a de Laval nozzle, a large fraction of the thermal energy of the gas is converted into kinetic energy. In the ideal case the resulting cooling is simply a function of the ratio of the areas of the nozzle's exit aperture and its throat \cite{sutton2016rocket}.
For a  monatomic  gas 
in the ideal case, the ratio of the temperature at the nozzle exit to the temperature inside the cell is shown as a function of this expansion ratio in Fig. \ref{fig:cooling}.
%
We also note that collisions in the CBGB beam after it exits the nozzle could drive additional cooling or heating not included in Fig. \ref{fig:cooling}.

The ideal case assumes sufficient densities that there is negligible heat transfer to the gas from the nozzle walls and  that the gas expands adiabatically throughout the nozzle. Achieving this limit requires high gas flows, as is the case for typical rocket engines and the high-flow CRESU experiments. 
Probing whether lower-flow CBGB sources can operate near this regime is explored in this experiment. 

%



\section{Experiment}

\begin{figure}[ht]
    \begin{center}
      \includegraphics[width=\linewidth]{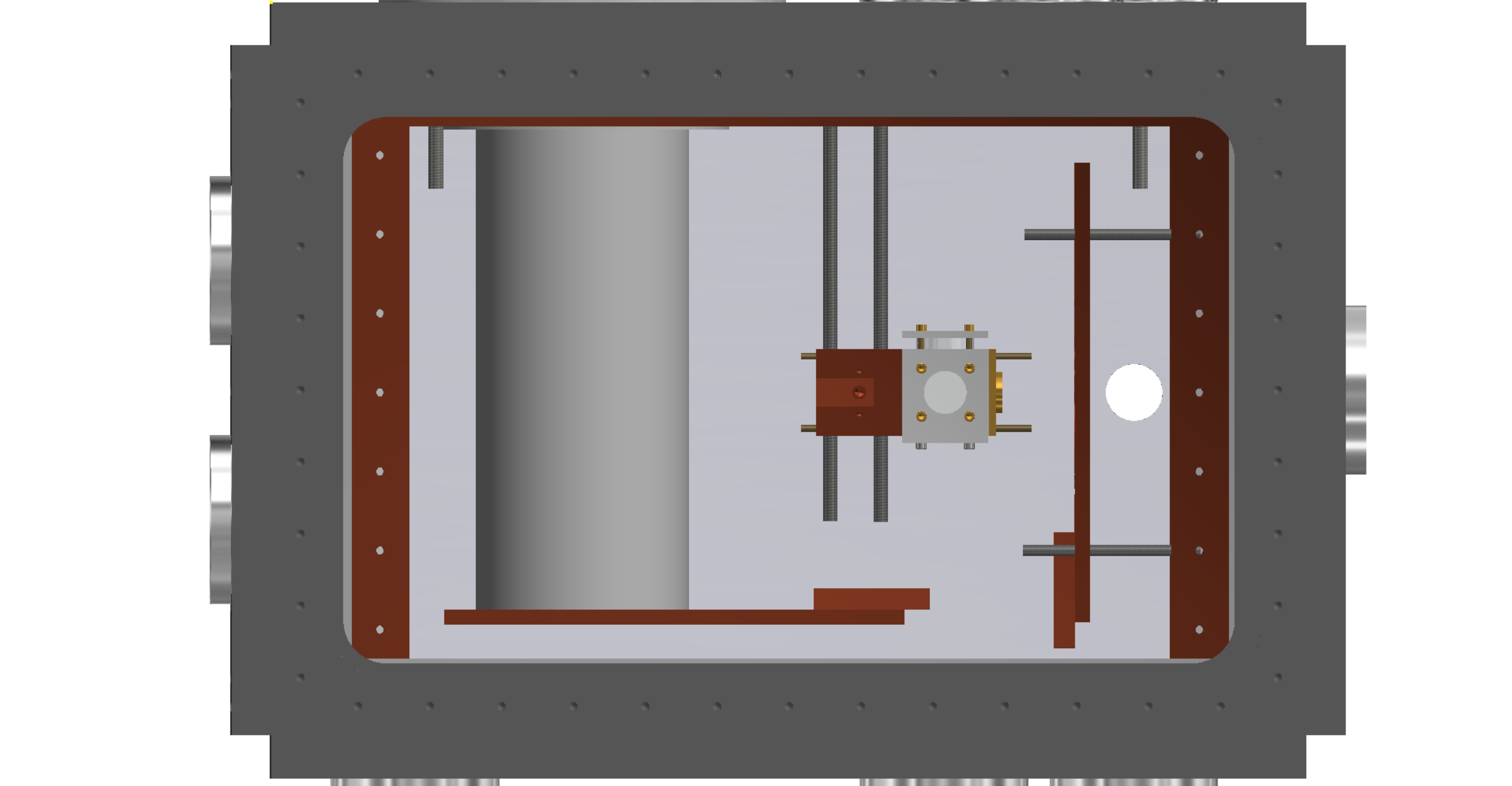}
    \caption{ \label{fig:apparatus} 
        Chamber schematic. The vacuum chamber is shown with one side open and one side of the radiation shield removed. The cell is at center right, cryogenic refrigerator to the left, and the two cryopumps are attached to the refrigerator bottom and to the right of the cell.  
                        }       
    \end{center}
\end{figure}

The cuboid vacuum chamber was machined from a solid 12" by 12" by 18" aluminum block, and sealed with aluminum plates, as shown in Fig. \ref{fig:apparatus}. Inside is a radiation shield built from copper and aluminum, cooled by the first stage of our cryogenic refrigerator (Cryomech PT415). Inside the radiation shield are the copper cell and cryopumps, cooled by the second stage of the refrigerator. 
A detailed description of the chamber can be found in reference \cite{Lancaster2018Thesis}.

Neon was used as the buffer gas for ease of cryopumping 
and its ability to produce beams of similar forward velocity and divergence as 4~K helium \cite{hutzler2012buffer}.

There are two cryopumping surfaces in the apparatus: one is directly anchored to the second stage of the cryocooler and the other suspended from the radiation shield and thermally linked to the first by copper braids (braids not shown in Fig. \ref{fig:apparatus}). The overall pumping area is 0.13~m$^2$, with both pumps residing below 5~K. With no other pumps running, these keep the chamber pressure below 10$^{-5}$~Torr at flow rates of up to 80~SCCM.
As seen in Fig. \ref{fig:apparatus}, the first cryopump sits at the bottom of the chamber, and the second cryopump is oriented vertically and sits 3.6 cm after the cell exit. It has a 2" diameter hole for the buffer gas beam to pass through. Better cryopumping could likely be achieved with a smaller hole; we used a large hole so as to not interfere with measurements of the beam angle.

\begin{figure}[ht]
    \begin{center}
      \includegraphics[width=\linewidth]{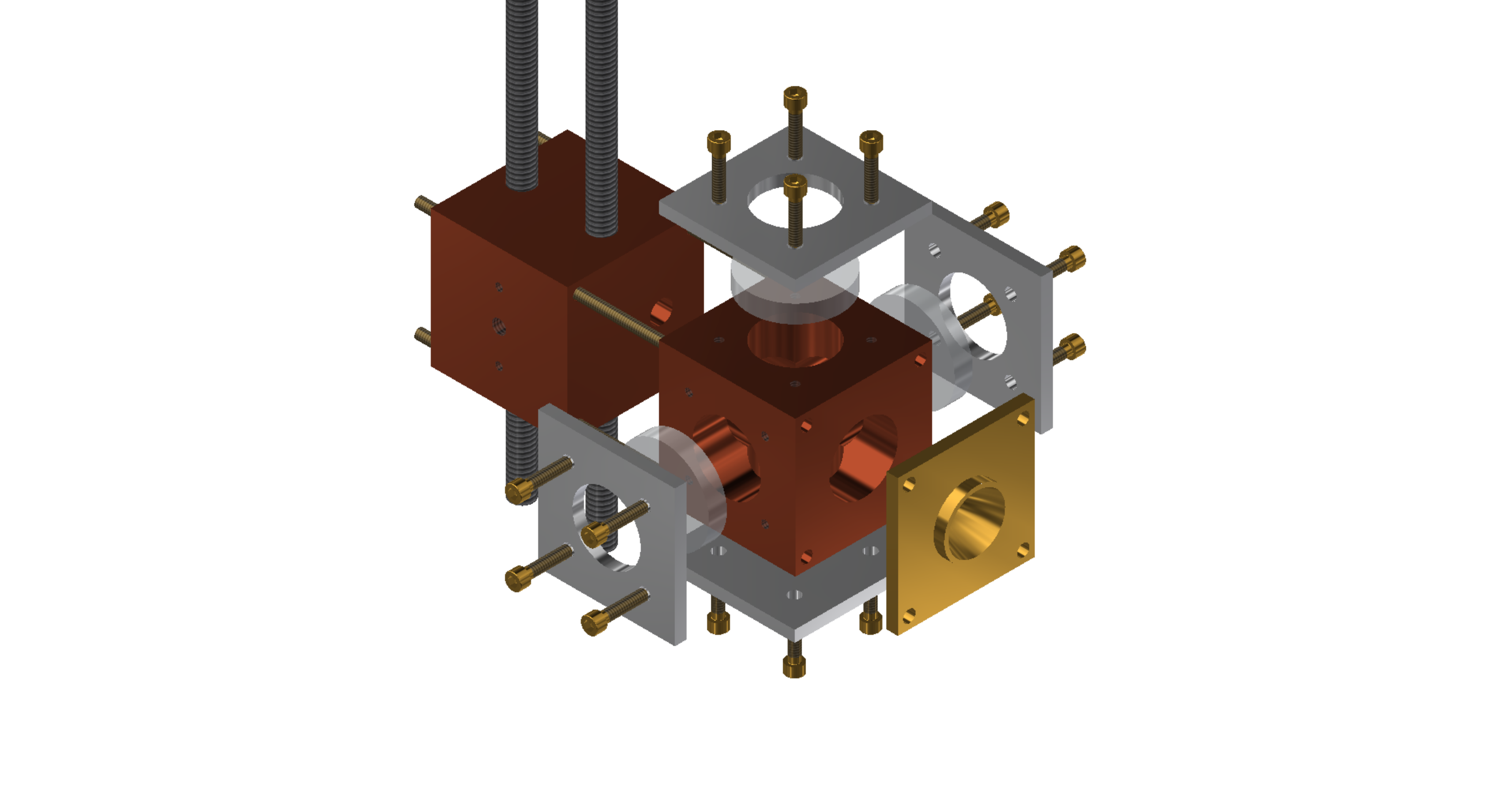}
    \caption{ \label{fig:cell} 
    Exploded schematic of the cell. 
    Threaded brass rods hold the copper cubes and the exit aperture together. Here the exit aperture is a brass nozzle.
 %
    	Not shown is the 1/4" OD copper tube attached to the back of the cell (at upper left) which supplies the buffer gas.
            }                   
    \end{center}
\end{figure}

The cell is comprised of two 1.5" copper cubes that are bolted together, as shown in Fig. \ref{fig:cell}. The first cube provides mechanical support and holds temperature sensors and heaters. The buffer gas is precooled by a $\sim 30$~cm long, 1/4" OD copper tube before flowing into the first cube.
The second cube has a 3/4" hole through each face, each sealed by a window or metal plate. This cube contains the ablation target, windows for probe beams and ablation beam, and the exit aperture.

Five ``simple'' apertures were tested, each consisting of a cylindrical hole through a 1/8" aluminum plate. The hole diameters tested were: 1/16", 1/8", 3/16", and 1/4", and a 3/16" beveled hole. The 3/16" beveled hole aperture was found to give the best results of these apertures: the beam produced by the 3/16" hole had higher optical depths than smaller apertures and better collimation than the larger hole or unbeveled 3/16" hole.
For the remainder of this paper, we use this aperture for comparison to the nozzles.

\begin{figure}[ht]
    \begin{center}
      \includegraphics[width=\linewidth]{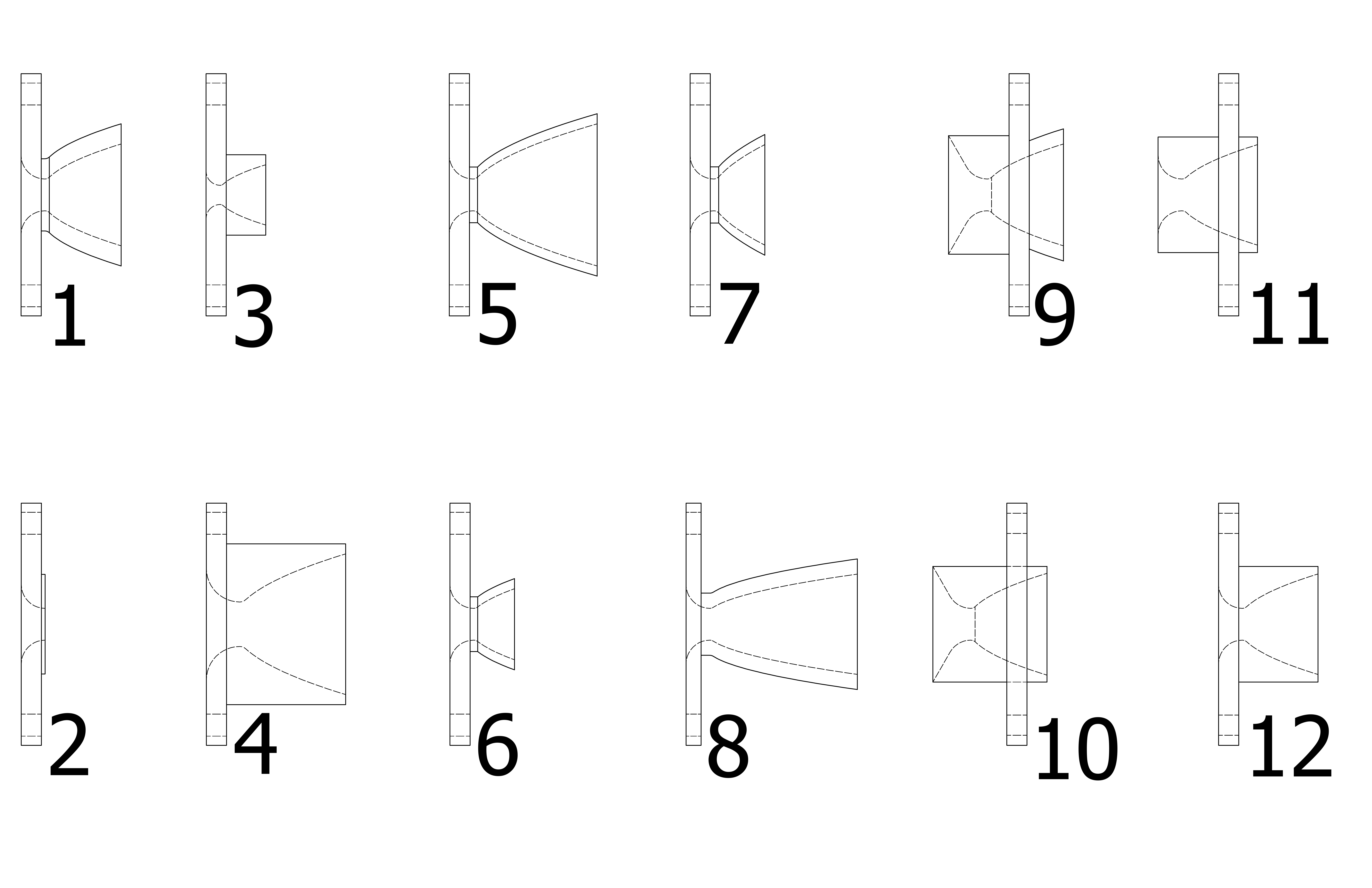}
    \caption{ \label{fig:nozzles} 
    Wireframe view of nozzles tested. 1--9 were 3-D printed from  nylon, and 10--12 were machined out of brass. All drawn to scale; the height of the baseplate is 1.5''.}        
    \end{center}
\end{figure}

Nine 3-D printed nylon nozzles were tested during the experiment, using a variety of designs based on textbook formulas  for bell-shaped nozzles following the ``Rao'' design \cite{rao1961recent, sutton2016rocket}.
These designs varied the three textbook parameters used to calculate nozzle shape: the throat diameter, the ratio of exit and throat areas, and the length. The designs tested are shown in Fig. \ref{fig:nozzles}, and detailed in reference  \cite{MackTaylorThesis2018}.

Following the nylon nozzles, three machined brass nozzles were tested, with designs which were based on the nylon nozzles that gave the best performance, using the same dimensions and scaling. 
These are also shown in Fig. \ref{fig:nozzles}.
%
The nozzle which gave the highest flux of atoms per solid angle was nozzle \#11. We will focus on the results of that nozzle for the remainder of the paper.

For convenience, the nozzles were tested with atomic beams rather than molecular beams. The experiment was run with both ytterbium and titanium atoms, to explore the effects of atomic mass on the beam properties. To produce these atoms, metal targets are ablated with a frequency-doubled 
Nd:YAG laser. The laser (Litron Nano) has a pulse energy of 20--60~mJ at a wavelength of 532 nm and a repetition rate of 10~Hz.  The ablation targets are at the bottom of the cell, and the ablation laser comes in through a window in the top of the chamber. 

The atoms are probed with a narrow-linewidth tunable blue diode laser (Toptica DL-100), with typical probe powers of $\lesssim 0.1$~mW. 
The probe laser goes through 
the chamber in three places to measure the in-cell atoms and the atomic beam.

The first probe beam is sent through the cell, which measures the in-cell atom population with absorption spectroscopy. 
All after-cell measurements were taken roughly 6 cm 
after the cell, after the beam has passed through the cryopump aperture. Based on prior work, this is far enough that collisions will have ``frozen out'' and the velocity distribution will no longer change  \cite{hutzler2011cryogenic}.
However, we note that at our very highest flow rates, calculations suggest that a small number of intra-beam collisions (of order 1) may occur after this point; it is unlikely that any such collisions will produce a significant change to the velocity distribution.

The second probe beam crosses the atomic beam $\sim6$~cm after the cell exit in the transverse direction and is double-passed through the atomic beam. This beam is measured via both absorption and fluorescence spectroscopy; the absorption is used to calibrate the fluorescence. 
The third probe beam is counterpropagating with respect to the atomic beam, and is measured via fluorescence collected at the same location as the transverse beam crosses the atomic beam. Because beams 2 and 3 use the same PMT to detect fluorescence they cannot be measured at the same time and are measured sequentially.



We note that the aperture in the cryopump limits the range of atomic beam angles that can reach the detection region, typically to 
$\lesssim 30^\circ$ with respect to the axis. 
This may artificially reduce our highest transverse temperature measurements.

\section{Experimental results}
\label{sec:ExpResults}

Measured in-cell temperatures for titanium (determined from  Doppler broadening) were typically $\sim20$~K for cells with apertures and nozzles, largely independent of the buffer gas flow rate.

\begin{figure}[ht]
    \begin{center}
      \includegraphics[width=\linewidth]{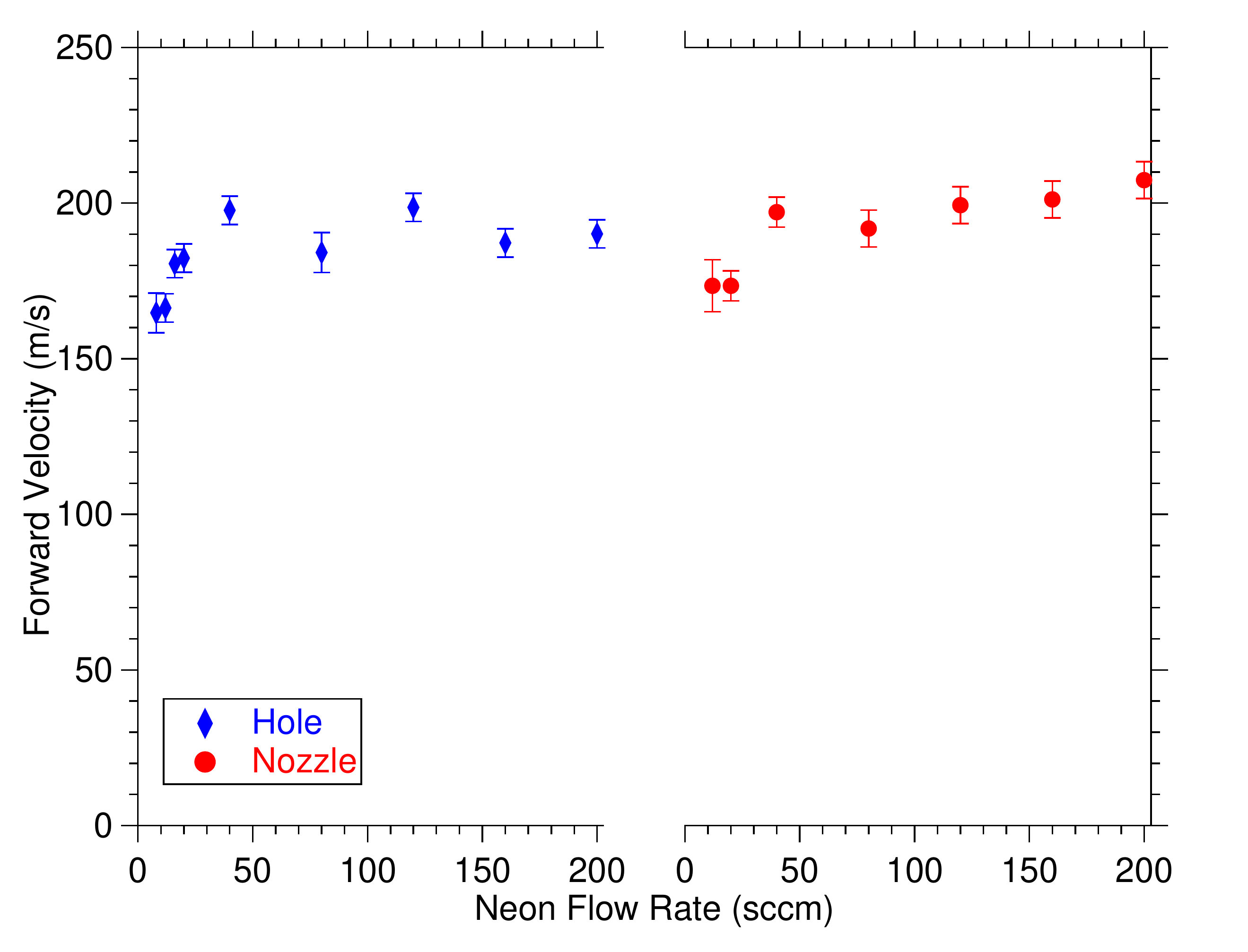}
    \caption{ \label{fig:Ti_vel} 
    Forward velocities vs neon flow rate for titanium ablation with a 3/16" beveled hole aperture (left) and a de Laval nozzle (right). 
                        }       
    \end{center}
\end{figure}

Similarly, the forward velocities seen in the beams (measured from the Doppler shift of the axial beam), show little difference between holes and de Laval nozzles, as seen in Fig. \ref{fig:Ti_vel}. Velocities are lower at low flow rates, and plateau at higher flow rates, as the flow enters the ``boosted'' regime, as seen in prior CBGB experiments \cite{barry2011bright,hutzler2011cryogenic}.

\begin{figure}[ht]
    \begin{center}
      \includegraphics[width=\linewidth]{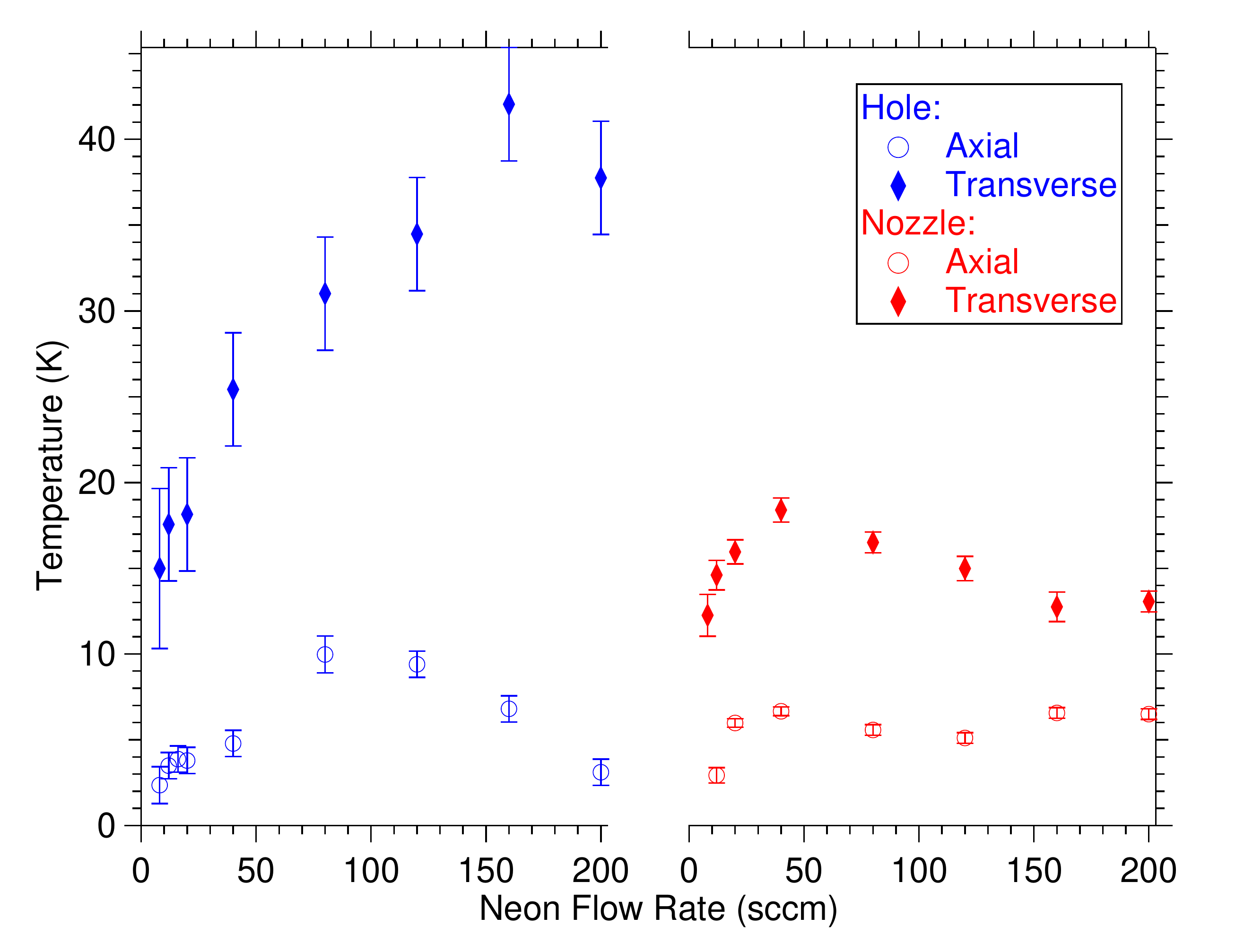}
    \caption{ \label{fig:Ti_temps} 
    Axial and transverse temperatures vs neon flow rate for a titanium atomic beam with a 3/16" beveled hole aperture (left) and a de Laval nozzle (right). 
                        }       
    \end{center}
\end{figure}

The measured temperatures in the beam show more complex behavior. Fig. \ref{fig:Ti_temps} shows measurements of the axial and transverse velocity distributions in the beam, as determined from Doppler spectroscopy.  We fit each to a Gaussian distribution and characterize it in terms of an equivalent temperature. 
The transverse velocity distribution typically shows very good agreement with a Gaussian. The axial velocity distribution often did not, especially at higher flows. We nevertheless use temperature to characterize the velocity spread in a simple manner. 

The de Laval nozzle provided a noticeable improvement in the transverse velocity spread. 
Both the hole and the nozzle have similar transverse velocity distributions at low flows (near the effusive regime). As the flow increases the transverse temperature continues to increase for the hole, but for the nozzle it ``turns over'' and decreases with flow over the 50--200 SCCM range. This leads to a significantly more collimated beam.

Axial temperatures also show differences between the hole and the nozzle, with no significant advantage to the nozzle, and slightly higher temperatures for the nozzle at certain flow rates. It is not understood why such low axial temperatures were measured at low flow rates for both the straight holes and the nozzles; little expansion cooling is expected in the effusive regime.


\begin{figure}[ht]
    \begin{center}
      \includegraphics[width=\linewidth]{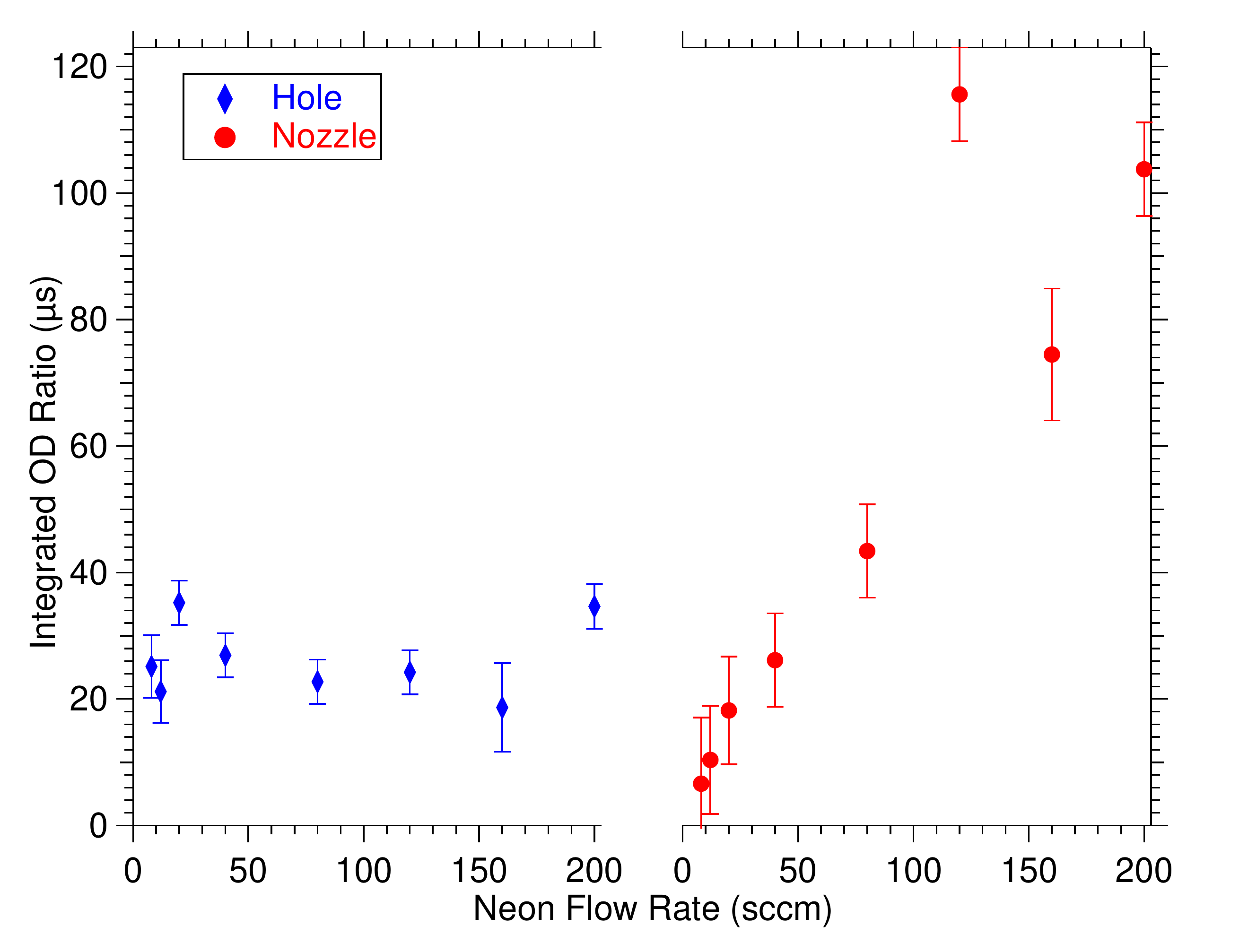}
    \caption{ \label{fig:Ti_OD} 
    Integrated OD ratios vs neon flow rate for titanium ablation with a 3/16" beveled hole aperture (left) and a de Laval nozzle (right). 
              }       
    \end{center}
\end{figure}

As seen in Figs. \ref{fig:Ti_vel} and \ref{fig:Ti_temps}, the nozzle gives a significantly more collimated beam at high buffer-gas flows. However, this is only advantageous if it produces a higher number of atoms per solid angle; otherwise the same collimation could be achieved with an aperture in the beam.
To measure this, we use the optical depth measured by the transverse probe after the cell (OD$_\mathrm{T}$).
%
 %
Far from the nozzle --- for constant forward velocity --- the rate of atoms arriving per unit area ($\frac{\dot{N}}{A}$) is proportional to the rate per unit solid angle ($\frac{\dot{N}}{\Omega}$).
%
We note $\mathrm{OD_T} = \int  n(z) \sigma(z) \ dz \approx n \sigma L$.
With the above assumptions, 
$L \propto v_{\mathrm{T}}$, where $v_{\mathrm{T}}$ is the transverse velocity spread of the atomic beam. Doppler broadening gives $\sigma \propto 1/v_{\mathrm{T}}$. Hence the on-axis density $n$ of our molecular beam 
is simply proportional to $\mathrm{OD_T}$.
$\frac{\dot{N}}{A}$ is the product of the density and the forward velocity so 
the total number of atoms per unit area per pulse is $\propto \int n \ dt \propto \int \mathrm{OD_T} \ dt$. 
%
To compensate for shot-to-shot fluctuations in ablation production,  the time-integrated OD is normalized by peak OD measured in the cell, which should be proportional to the number of atoms produced. This is shown in  Fig \ref{fig:Ti_OD}.
Unfortunately the OD in the cell at early times is too large to measure directly; we extrapolate the OD measured at later times to the time of ablation assuming exponential decay (which is typically a good approximation at times where the OD can be observed).


As seen in Fig. \ref{fig:Ti_OD}, at low gas flows the hole outperforms the nozzle in terms of numbers of Ti atoms per solid angle per pulse. The hole produces roughly constant flux per solid angle for the flow range explored, but the nozzle continues to improve with increasing flow. 
The de Laval nozzle produces 
a greater flux of atoms per solid angle per ablation pulse than the hole for flow rates above 40 SCCM, as expected from the improved beam collimation observed in Figs. \ref{fig:Ti_vel} and \ref{fig:Ti_temps}. 
%
We note that in our experiments, higher gas flows generally produced higher atomic densities in the cell, so if Fig. \ref{fig:Ti_OD} was not normalized to the in-cell OD the improvement seen at high flows would be even larger.


Measurements for Yb exhibit similar behavior to Ti, as shown in detail in section \ref{sec:Ybfigs} and summarized here. The forward velocity behaves qualitatively similarly, but with a slightly lower asymptote of $\sim 175$~m/s.

Ytterbium's axial and transverse temperatures show qualitatively similar behavior to titanium. Quantitatively, at high flows the nozzle data shows higher transverse temperatures for Yb than Ti, with a transverse temperature of $22 \pm 2$~K at high neon flows. 

Integrated OD ratios show qualitatively similar behavior to titanium, with the highest number of atoms per solid angle per pulse obtained with the nozzle operated at high buffer gas flow.

\section{Mass dependence}

The ``temperature'' used to characterize the transverse velocity spread is higher for Yb than Ti. However, due to Yb's larger mass, the Yb temperatures corresponds to a narrower transverse velocity spread at high flows: a FWHM of $76\pm4$~m/s for Yb compared to $107\pm5$~m/s for Ti. 

\begin{figure}[ht]
    \begin{center}
      \includegraphics[width=.55\linewidth]{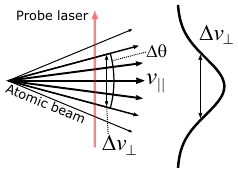}
    \caption{
    \label{fig:AngularSpread} 
    An illustration of gas flow. Because the axial probe has a narrow spread, it essentially samples a single ``flow line'' and measures the axial gas temperature directly. The transverse probe samples across all flow lines and thus the measured velocity distribution is a convolution of flow and local thermal motion.
	}     
    \end{center}
\end{figure}

We model the velocity distribution as an average macroscopic flow convolved with local thermal motion, with the flow depicted schematically in Fig \ref{fig:AngularSpread}.
For the forward velocity distribution the \emph{on-axis} macroscopic flow is a single mean velocity; the Gaussian fit to the distribution of velocities around this mean gives the temperature directly.

\begin{figure}[ht]
    \begin{center}
      \includegraphics[width=\linewidth]{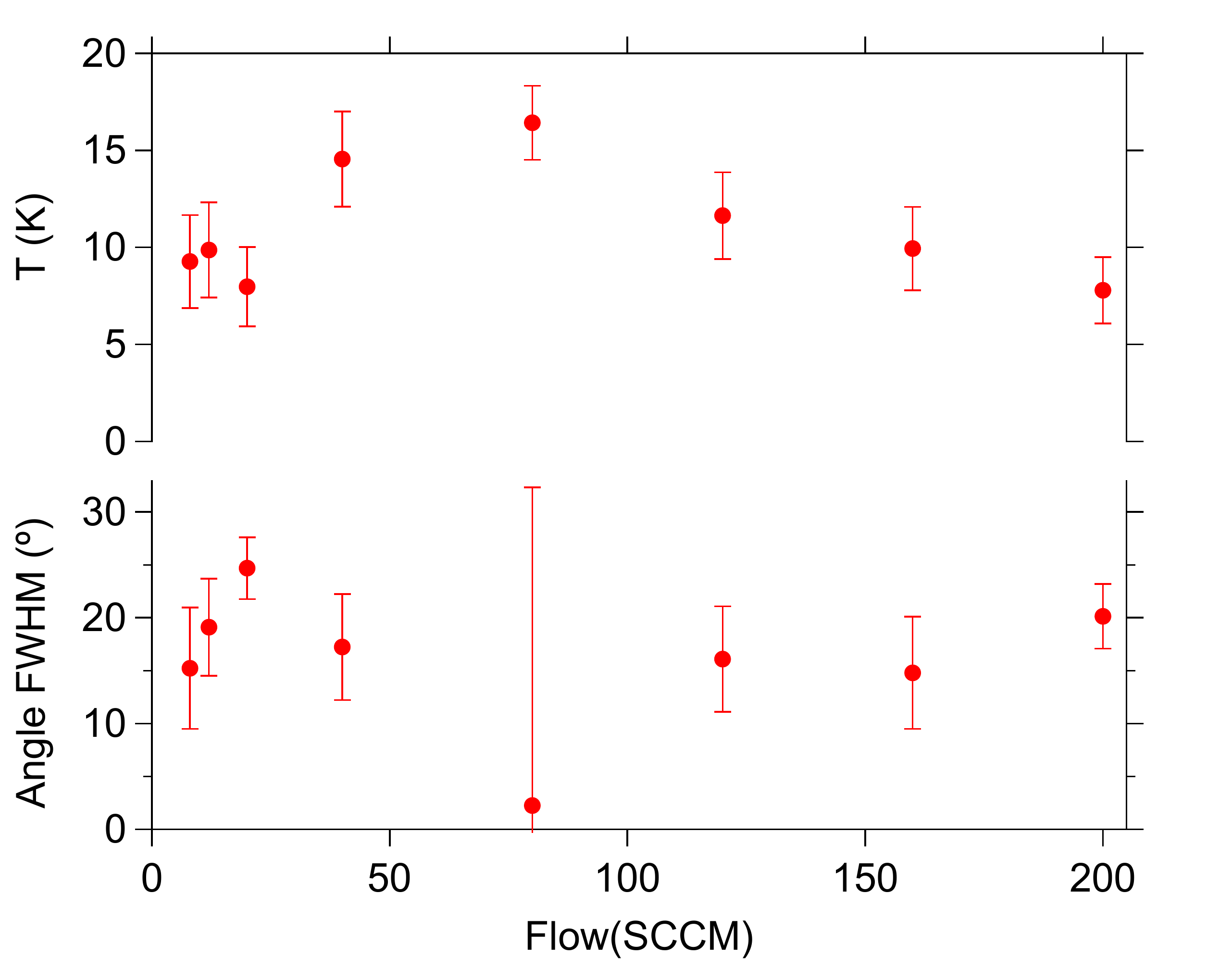}
    \caption{
    \label{fig:TransverseTAndAngle} 
    The extracted flow angle (assuming a constant forward velocity of 200~m/s) and transverse temperature of the entrained atoms, as determined from the Ti and Yb data.
	}     
    \end{center}
\end{figure}

For the transverse velocity distribution, the situation is more complicated: we sample flow over the entire transverse spatial distribution, as seen in Fig \ref{fig:AngularSpread}.
We assume the local velocity distribution to be a Maxwell-Boltzmann distribution; for simplicity, we also model the angular flow pattern as a Gaussian distribution. Thus the measured velocity distribution is a convolution of these two distributions, with a width $\sigma_{\mathrm{measured}} = \sqrt{\sigma_{\mathrm{flow}}^2 + \sigma_{\mathrm{T}}^2}$.
%
We can thus combine our measurements for Yb and Ti to extract both the flow characteristics and the gas temperature, assuming that these parameters are independent of the entrained species. At our highest flows, we find a transverse temperature of $8 \pm 2$~K, and a transverse flow velocity FWHM of $60 \pm 14$~m/s. The results as a function of flow are shown in Fig \ref{fig:TransverseTAndAngle}.

To within our signal-to-noise, we cannot discern a trend in the flow angle with buffer gas flow. For flows $\geq80$~SCCM, the data indicates that the temperature of the gas decreases with increasing flows.

\section{Discussion}

The converging-diverging nozzle outperforms a simple hole in both a reduced angular spread and a greater number of atoms per solid angle per ablation pulse. Achieving these improvements requires a higher buffer gas flux, which comes with greater requirements on cryopumping and differential pumping to achieve good vacuum in the beam region. Running at high flows also puts the beam fully into the ``boosted'' regime, which may be disadvantageous for experiments seeking to minimize the forward velocity of the beam.

\section*{Acknowledgements}
We gratefully acknowledge helpful discussions with John M. Doyle.  We gratefully acknowledge assistance in the construction of our apparatus, cryogenic cell, and nozzles from Wade J. Cline; Carl D. Davidson, Jr.; and Jake Holland.
This material is based upon work supported by the AFOSR under Grant No. FA9550-16-1-0117.

\bibliography{Nozzle2018}

\pagebreak

\section{Appendix: Yb data}
\label{sec:Ybfigs}

\begin{figure}[ht]
    \begin{center}
      \includegraphics[width=\linewidth]{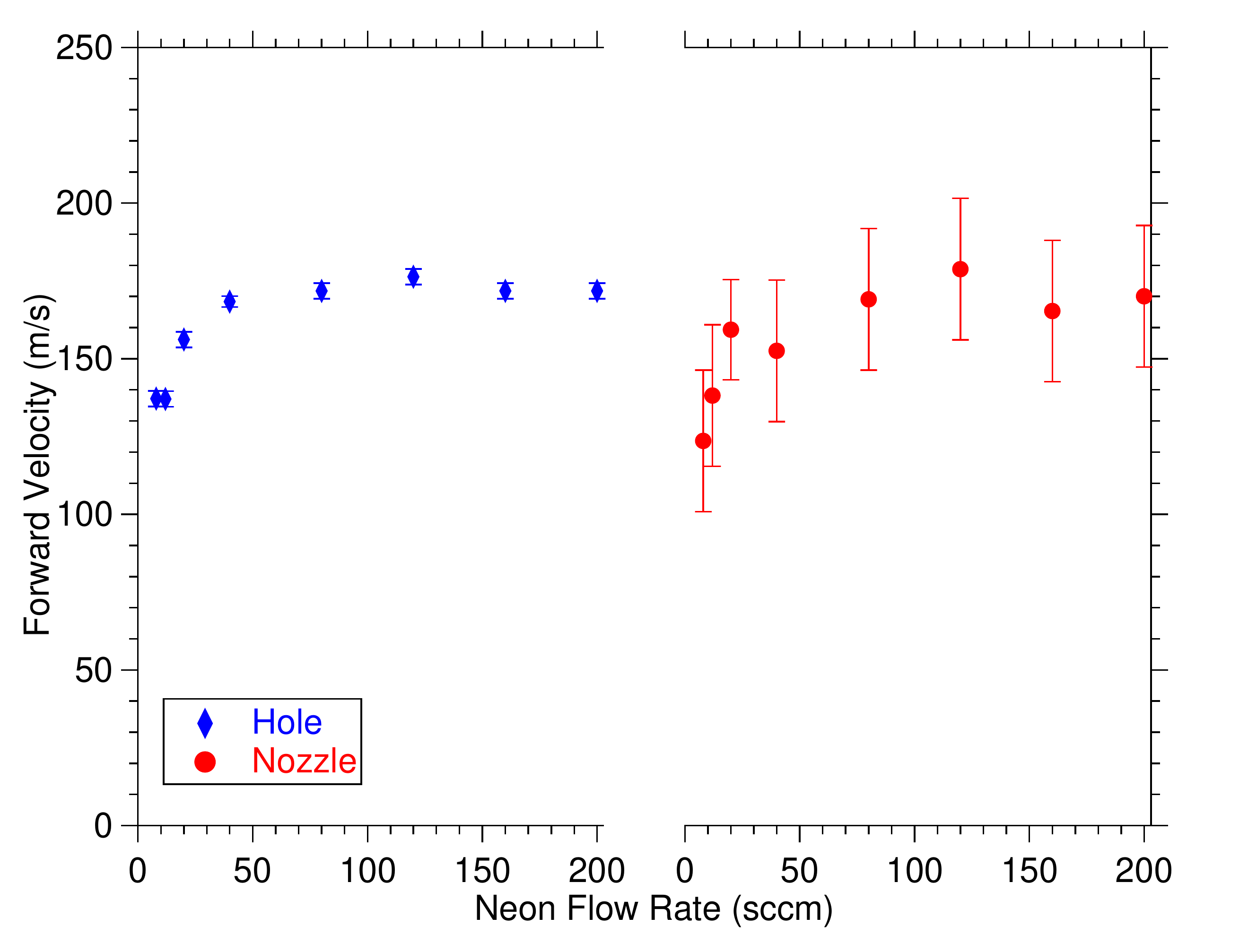}
    \caption{ \label{fig:Yb_vel} 
    Forward velocities vs neon flow rate for ytterbium ablation with a 3/16" beveled hole aperture (left) and a de Laval nozzle (right). 
                        }       
    \end{center}
\end{figure}

\begin{figure}[ht]
    \begin{center}
      \includegraphics[width=\linewidth]{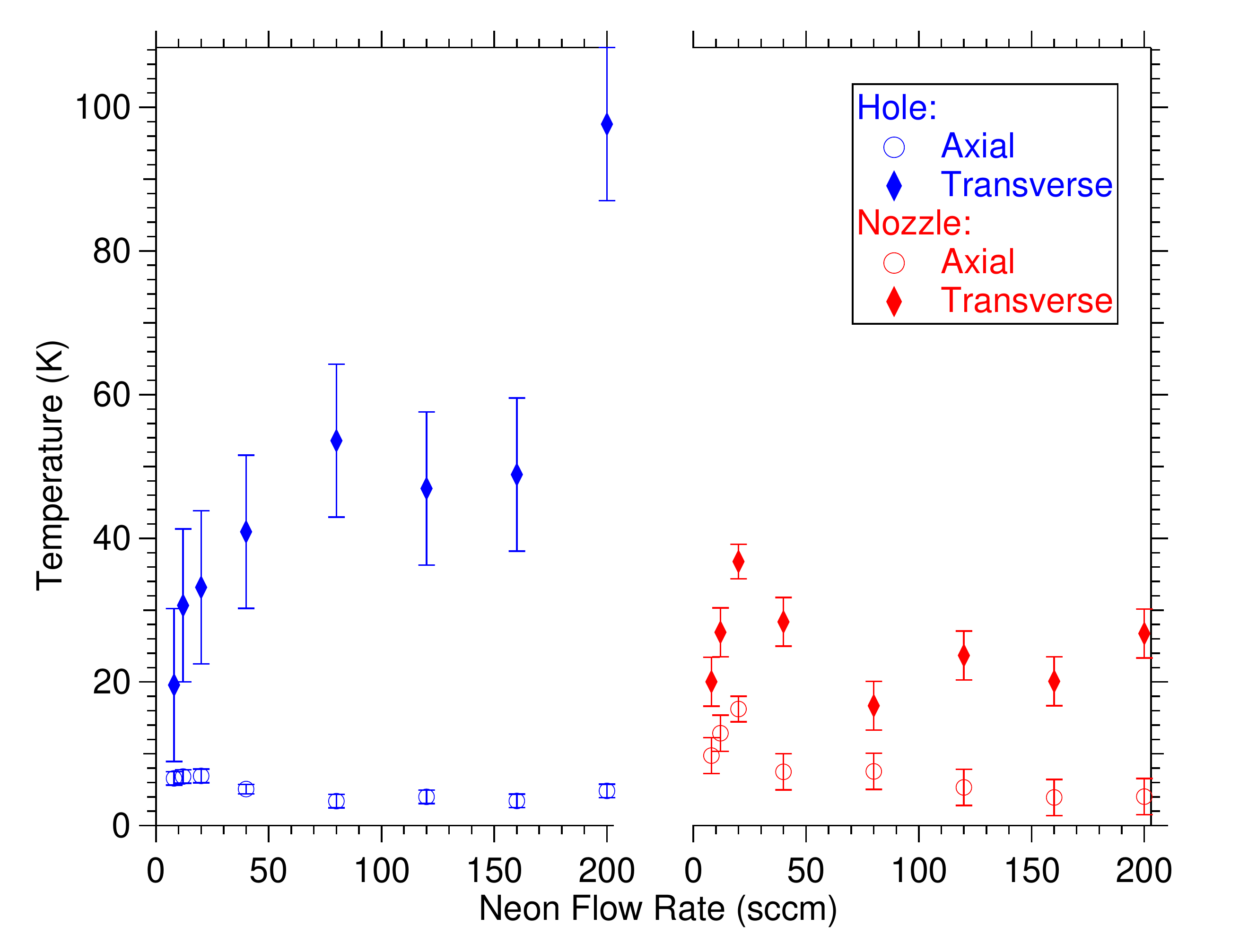}
    \caption{ \label{fig:Yb_temps} 
    Axial and transverse temperatures vs neon flow rate for ytterbium ablation with a 3/16" beveled hole aperture (left) and a de Laval nozzle (right). 
                        }       
    \end{center}
\end{figure}

\begin{figure}[ht]
    \begin{center}
      \includegraphics[width=\linewidth]{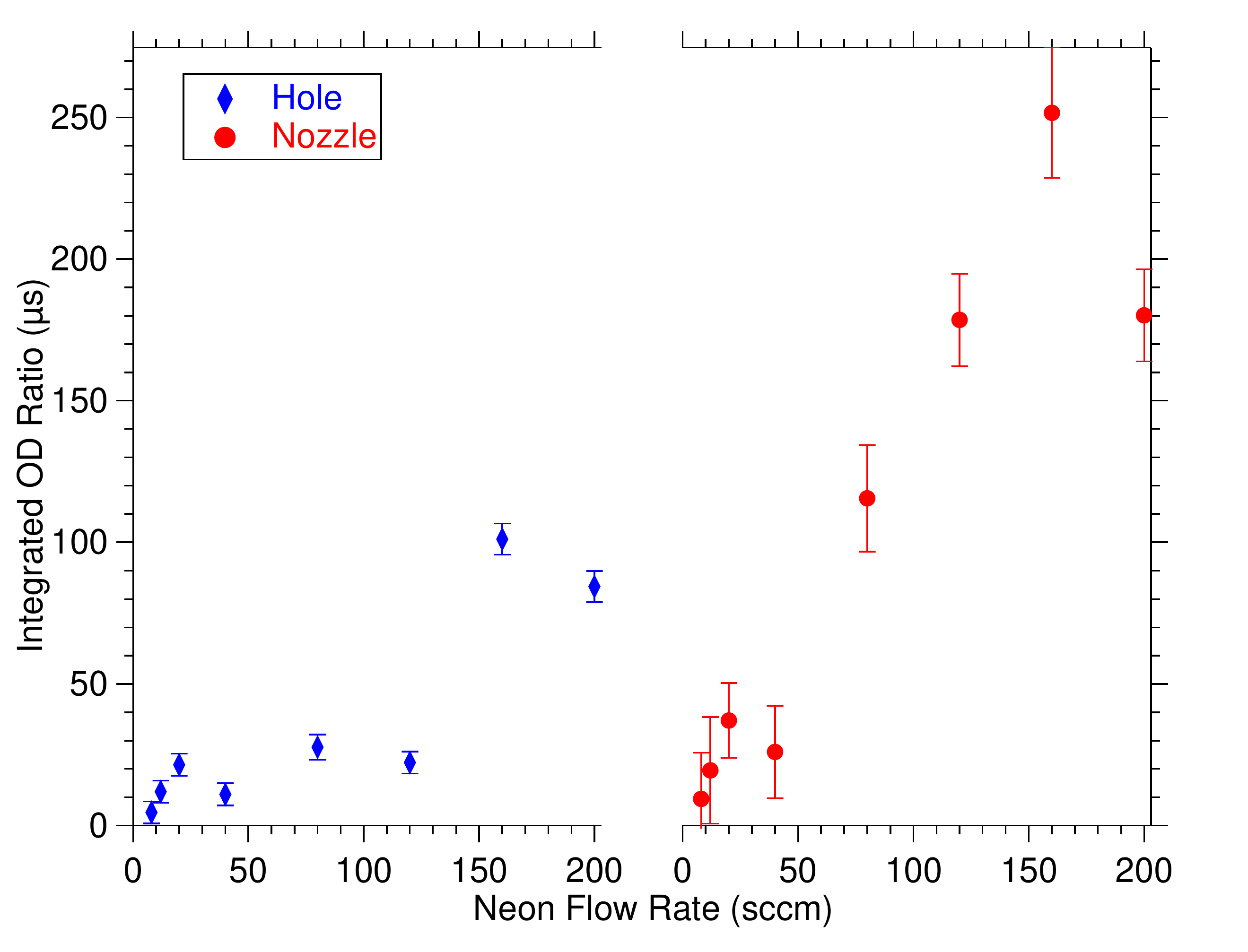}
    \caption{ \label{fig:Yb_OD} 
    Integrated OD ratios vs neon flow rate for ytterbium ablation with a 3/16" beveled hole aperture (left) and a de Laval nozzle (right). 
                        }       
    \end{center}
\end{figure}

We have produced and measured a CBGB of Yb atoms using the same simple hole aperture and de Laval nozzle that produced the best results for Ti. Due to the extremely large fluxes of atoms and high optical depths observed (and lack of titanium's easily frequency-resolved low-natural-abundance isotopes) the Yb data is noisier than the Ti data. 
The measured on-axis axial velocities are presented in Fig. \ref{fig:Yb_vel}. The measured axial and transverse velocity distributions are shown in Fig. \ref{fig:Yb_temps}; as for Ti atoms, the velocity distribution is characterized by an equivalent temperature.
The integrated OD ratio, as described in Section \ref{sec:ExpResults} is presented in figure \ref{fig:Yb_OD}.

\end{document}